# Decision-making under uncertainty – a quantum value operator approach

Lizhi Xin[1] and Houwen Xin[2, *]

[1]Building 59, 96 Jinzhai Road, Hefei, Anhui, P. R. China
[2]Department of Chemical physics USTC, Hefei, Anhui, P. R. China
[*]hxin@ustc.edu.cn

**ABSTRACT**

Decision-making under uncertainty is the unification of people's subjective beliefs and the objective world. We propose a quantum expected value theory for decision-making under uncertainty. Quantum density operator as value operator is proposed to simulate people's subjective beliefs. Value operator guides people to choose corresponding actions based on their subjective beliefs through objective world. The value operator can be constructed from quantum gates and logic operations as a quantum decision tree. The genetic programming is applied to optimize and auto-generate quantum decision trees.

**Introduction**

Decision-making can be viewed as a two-phase process: evaluation process and selection process. Classic decision theory holds that rational-economic person knows the utility function as well as probability distribution through evaluation process then maximizes utility by an optimal selection[1-3]. Information completeness and selection consistence are required for expected utility theory, however in the real world information is rarely complete and consistency of selection cannot be guaranteed due to complexity and people show irrational behaviors[4-7] which cannot be explained by classic decision theory.

Classic decision theory is a "black box"; people do not know what really happens inside the box. Scientists are trying to apply quantum theory to reveal how people make decisions[8-10]. We believe that the results of decisions are the unity of subjective beliefs and objective facts, and the value of observed results is the bridge between those two different worlds. The decision's "black box" can be opened through the bridge of the value.

**Quantum expected value decision theory**

Usually people subjectively choose an action $a_i \in \{a_1, \cdots, a_m\}$ where nature's objective state is in $q_j \in \{q_1, \cdots, q_n\}$ when decisions were made, and the result of the decision value matrix $v_{ij}$ depends on both the state of the nature and choice of brain shown in table 1. Natural state describes the objective world; we hypothesize that an uncertain natural state can be represented by superposition of all possible states in terms of the Hilbert state space[11] as in (1). Strategy state describes the subjective world; we also hypothesize that undecided decision state can be represented by superposition of all possible actions as in (2). Usually the information of decision-making under uncertainty is incomplete, the result of decision can be represented by a mixed state's density operator as a value operator as in (3). Value operator is a sum of projection operators which projects a person's beliefs onto an action of choice based on nature states. Quantum expected value can be represented as in (5). The uncertainty of observed value can be represented by Shannon information entropy[12] as in (6).

| State \ Action | $q_1$ | $\cdots$ | $q_j$ | $\cdots$ | $q_n$ |
|---|---|---|---|---|---|
| $a_1$ | $v_{11}$ | | $\cdots$ | | $v_{1n}$ |
| $\vdots$ | | $\ddots$ | | | |
| $a_i$ | $\vdots$ | | $v_{ij}$ | | $\vdots$ |
| $\vdots$ | | | | $\ddots$ | |
| $a_m$ | $v_{m1}$ | | $\cdots$ | | $v_{mn}$ |

**Table 1** State-action-value decision table

$$|\psi\rangle = \sum_j c_j |q_j\rangle, \quad \sum_j |c_j|^2 = 1 \tag{1}$$

$$|S\rangle = \sum_i \mu_i |a_i\rangle, \quad \sum_i |\mu_i|^2 = 1 \tag{2}$$

$$\hat{V} = \sum_i p_i |a_i\rangle\langle a_i|, \quad \sum_i p_i = 1 \tag{3}$$

$$<\hat{V}> = \langle\psi|\hat{V}|\psi\rangle = \sum_k c_k^* \langle q_k| \sum_i p_i |a_i\rangle\langle a_i| \sum_j c_j |q_j\rangle \tag{4}$$

$\langle q_k | q_j \rangle = \delta_{jk}$, $|q_j\rangle$ is a set of quantum orthonormal basis. We have:

$$<\hat{V}> = \sum_i p_i \sum_j |c_j|^2 |\langle a_i | q_j \rangle|^2 = \sum_i p_i \sum_j \omega_j v_{ij} \tag{5}$$

$$\Delta V = \Delta V_{subjective} + \Delta V_{objective} = -\left(\sum_i p_i \log p_i + \sum_j \omega_j \log \omega_j\right) \tag{6}$$

Where $p_i = |\mu_i|^2$ is a person's subjective belief in choosing an action $a_i$; $\omega_j = |c_j|^2$ is the objective frequency at which natural state is in $q_j$; value matrix $v_{ij} = |\langle a_i | q_j \rangle|^2$ is the results of decision when a person chooses an action $a_i$ based on natural state $q_j$. Quantum expected value decision theory suggests that a subjective and objective unified result is obtained through people's beliefs which are based on natural states.

The decision process of a person can be simulated by the continuous evolution of the value operator according to the information as in Liouville equation (7). Eigen equation of information energy operator is represented as in (10). A final decision is equivalent to a Von Neumann projection measurement[13] performed on the strategy state by the brain which chooses an action $a_i$ with probability $p_i$ as in equation (11).

$$i \frac{d}{dt} \hat{V}(t) = [\hat{H}(f(t)), \hat{V}(t)] \tag{7}$$

$$f(t) = \{A_i(t), i = 1, 2, \cdots, N\} \tag{8}$$

$$A_i(t) = a_j \in \{a_1, \cdots, a_m\} \tag{9}$$

$$\hat{H}|q_j\rangle = E_j|q_j\rangle \tag{10}$$

$$D: \boxed{A} \xrightarrow{projection} |S\rangle_{p_i} \to |a_i\rangle \tag{11}$$

Where $\hat{H}$ the information energy operator is determined by all N decision makers as in (8); (9) means that the ith decision maker $A_i$ takes the jth action $a_j$ at time t; $E_j$ is all the information that decision makers can get when natural state is in $q_j$. Information is the essence of people's subjective beliefs just like energy is the essence of the objective world. Valuable information can reduce uncertainty.

As an example, we can represent the state of future market in terms of the Hilbert state space as in (12), Hilbert strategy space is applied to represent trader's strategies as in (13), quantum density operator as value operator which projects a trader's beliefs onto an action of buying or selling a security as in (14).

$$|\psi\rangle = c_1|q_1\rangle + c_2|q_2\rangle \tag{12}$$

$$|S\rangle = \mu_1|a_1\rangle + \mu_2|a_2\rangle \tag{13}$$

$$\hat{V} = p_1|a_1\rangle\langle a_1| + p_2|a_2\rangle\langle a_2| \tag{14}$$

Where $|q_1\rangle$ indicates a state in which the market is rising and $|q_2\rangle$ indicates a state in which the market is falling; $|a_1\rangle$ represents trader's action to buy and $|a_2\rangle$ represents trader's action to sell; $p_1$ represents the subjective probability which a trader choose to buy and $p_2$ represents the subjective probability which a trader choose to sell.

Quantum expected value can be represented as in (15). The uncertainty of observed value can be represented by Shannon information entropy as in (16).

$$<\hat{V}> = \sum_{i=1,2} p_i \sum_{j=1,2} \omega_j v_{ij} = (2p-1)(2\omega-1)v \tag{15}$$

$$\begin{aligned}\Delta V = &-\left(\sum_{i=1,2} p_i \log p_i + \sum_{j=1,2} \omega_j \log \omega_j\right)\\ =& -\{[p \log p + (1-p)\log(1-p)]\\ &+ [\omega \log \omega + (1-\omega)\log(1-\omega)]\}\end{aligned} \tag{16}$$

Where p represents the subjective probability which a trader choose to buy; $\omega$ is the objective frequency at which market state is rising. v is the absolute value of the difference between the open price and the close price. Quantum expected value is between –v and +v while uncertainty of observed value is between 0 and 2.

There are three different situations:

1. Complete certainty
   a) $(p = \omega = 1 \,||\, p = \omega = 0) \to (<\hat{V}> = v,\ \Delta V = 0)$
   b) $(p = 1,\ \omega = 0 \,||\, p = 0,\ \omega = 1) \to (<\hat{V}> = -v,\ \Delta V = 0)$
2. Complete uncertainty
   $\left(p = \omega = \frac{1}{2}\right) \to (<\hat{V}> = 0,\ \Delta V = 2)$
3. Real market
   $(0 \leq p \leq 1,\ 0 \leq \omega \leq 1) \to (-v \leq <\hat{V}> \leq v,\ 0 \leq \Delta V \leq 2)$

**Quantum decision tree (qDT) is used to simulate people's subjective beliefs**

Master equation (7) is difficult to solve accurately, but we can approximately obtain a value operator by constructing a qDT which composed of quantum gates[14] and logic operations. The qDT composes of different nodes and branches. There are two types of nodes, non-leaf nodes and leaf nodes. The non-leaf nodes are composed of the operation set F as in (17); the leaf nodes are composed of the data set T as in (18) and (19). The construction process of a qDT is to randomly select a logic symbol from the operation set F as the root of qDT, and then grows corresponding branches according to the nature of the operation symbol and so on until a leaf node is reached.

$$F = \{+(\text{ADD}), \quad *(\text{MULTIPLY}), \quad //(\text{OR})\} \tag{17}$$

$$T = \{H, X, Y, Z, S, D, T, I\} \tag{18}$$

$$\left\{\begin{array}{l}H = \frac{1}{\sqrt{2}}\begin{bmatrix}1 & 1\\1 & -1\end{bmatrix}\ X = \begin{bmatrix}0 & 1\\1 & 0\end{bmatrix}\ Y = \begin{bmatrix}0 & -i\\i & 0\end{bmatrix}\ Z = \begin{bmatrix}1 & 0\\0 & -1\end{bmatrix}\\ S = \begin{bmatrix}1 & 0\\0 & i\end{bmatrix}\ D = \begin{bmatrix}0 & 1\\-1 & 0\end{bmatrix}\ T = \begin{bmatrix}1 & 0\\0 & e^{i\pi/4}\end{bmatrix}\ I = \begin{bmatrix}1 & 0\\0 & 1\end{bmatrix}\end{array}\right\} \tag{19}$$

The qDT of a value operator is a 2x2 matrix, and the value matrix needs to be diagonalized first and then normalized to get probability $p_1$ and $p_2$ as in (20), (21) and (22).

$$\widehat{V} = \begin{bmatrix} v_{11} & v_{12} \\ v_{21} & v_{22} \end{bmatrix} \xrightarrow{\text{diagonalization}} \begin{bmatrix} \lambda_1 & 0 \\ 0 & \lambda_2 \end{bmatrix} \xrightarrow{\text{normalization}} \begin{bmatrix} p_1 & 0 \\ 0 & p_2 \end{bmatrix} = p_1 |a_1\rangle\langle a_1| + p_2 |a_2\rangle\langle a_2| \tag{20}$$

$$|a_1\rangle = \begin{bmatrix} 1 \\ 0 \end{bmatrix}, |a_2\rangle = \begin{bmatrix} 0 \\ 1 \end{bmatrix}; \ |a_1\rangle\langle a_1| = \begin{bmatrix} 1 & 0 \\ 0 & 0 \end{bmatrix}, |a_2\rangle\langle a_2| = \begin{bmatrix} 0 & 0 \\ 0 & 1 \end{bmatrix} \tag{21}$$

$$\begin{vmatrix} v_{11} - \lambda & v_{12} \\ v_{21} & v_{22} - \lambda \end{vmatrix} = 0 \tag{22}$$

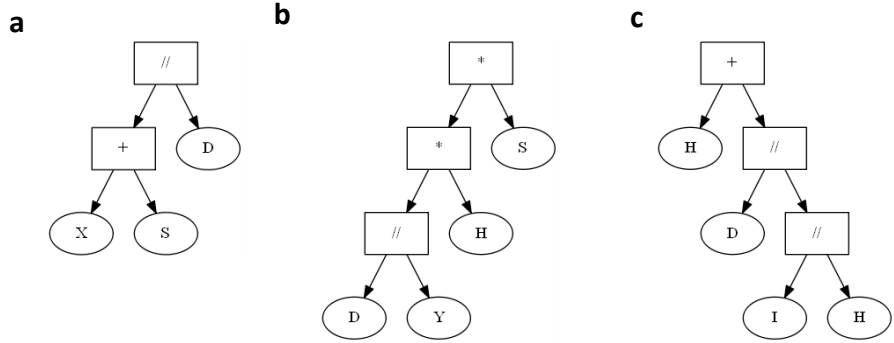

**Figure 1 A trader's subjective beliefs can be simulated by quantum decision trees.**
**(a)** $\text{qDT} = \big((X + S) // D\big)$ **(b)** $\text{qDT} = \Big(\big((D // Y) * H\big) * S\Big)$ **(c)** $\text{qDT} = \Big(H + \big(D // (I // H)\big)\Big)$

A trader's subjectively beliefs can be simulated by qDTs. For example, there are a few strategies with different subjective beliefs that the trader took on trading securities as shown in Figure 1.

- $\text{qDT} = \big((X + S) // D\big)$

1) $(X + S) \rightarrow \widehat{V} = 0.69|a_1\rangle\langle a_1| + 0.31|a_2\rangle\langle a_2|$ (69% belief to buy, 31% belief to sell)

2) $D \rightarrow \widehat{V} = 0.5|a_1\rangle\langle a_1| + 0.5|a_2\rangle\langle a_2|$ (50% belief to buy, 50% belief to sell)

- $\text{qDT} = \Big(\big((D // Y) * H\big) * S\Big)$

1) $\big((D * H) * S\big) \rightarrow \widehat{V} = 0.55|a_1\rangle\langle a_1| + 0.45|a_2\rangle\langle a_2|$ (55% belief to buy, 45% belief to sell)

2) $\big((Y * H) * S\big) \rightarrow \widehat{V} = 0.45|a_1\rangle\langle a_1| + 0.55|a_2\rangle\langle a_2|$ (45% belief to buy, 55% belief to sell)

- $\text{qDT} = \Big(H + \big(D // (I // H)\big)\Big)$

1) $(H + D) \rightarrow \widehat{V} = 0.5|a_1\rangle\langle a_1| + 0.5|a_2\rangle\langle a_2|$ (50% belief to buy, 50% belief to sell)

2) $(H + H) \rightarrow \widehat{V} = 0.5|a_1\rangle\langle a_1| + 0.5|a_2\rangle\langle a_2|$ (50% belief to buy, 50% belief to sell)

3) $(H + I) \rightarrow \widehat{V} = |a_1\rangle\langle a_1|$ (100% belief to buy)

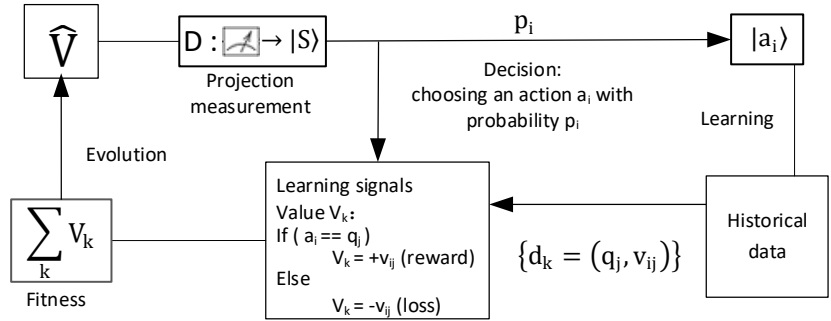

**Figure 2 The evolution of the quantum value operator**

The qDT can be optimized by the genetic programming (GP) [15, 16] as shown in Figure 2. A tree structure is used for encoding by GP, which is particularly appropriate to solve hierarchical and structured complex problems. An optimization problem mainly includes the selection of evaluation function and the acquisition of optimal solution. The evaluation function of qDT is a fitness function (24) based on observed value $V_k$ (23), and the optimal solution is obtained through continuous evolution by using selection, crossover and mutation as in (25); $A_i = |a_i\rangle\langle a_i|$ is Von Neumann's projection operator and $\max_{loss}\{V_k\}$ is the biggest loss of trading. If market is rising and a trader chooses to buy, the value $V_k$ is positive and the trader would make a profit, otherwise value $V_k$ is negative and the trader would lose money; if market is falling and a trader chooses to sell, the value $V_k$ is positive and the trader would make a profit, otherwise value $V_k$ is negative and the trader would lose money. The purpose of GP iterative evolution is to find a satisfactory qDT through learning historical data.

$$V_k = \langle q_j|A_i|q_j\rangle = |\langle a_i|q_j\rangle|^2 = v_{ij} = \begin{cases} -v, & i \neq j \\ v, & i = j \end{cases} \tag{23}$$

$$f_{fitness} = \left(\sum_{k=0}^{N} V_k\right) \Big/ \max_{loss}\{V_k\} \tag{24}$$

$$qDT \xrightarrow{evolution} \underset{qDT \in (F \cup T)}{\arg\max} (f_{fitness}) \tag{25}$$

**Algorithm: the evolution of quantum decision tree**

*Input*:
- Training data set $\{d_k = (q_j, v_{ij})\}$ which includes N samples, each sample consists of natural state and value.
- Setting
1) Operation set $F = \{+, *, //\}$
2) Data set $T = \{H, X, Y, Z, S, D, T, I\}$, eight basic quantum gates
3) Crossover probability = 70%; Mutation probability = 5%.

*Initialization*:
- Population: randomly create 100 ~ 500 qDTs.

*Evolution*:
- for k = 0 to N or a satisfactory generation is evolved:
a) Calculate fitness for each qDT.
b) According to the quality of fitness:
   i. Selection: selecting parent qDTs.
   ii. Crossover: generate a new offspring using the roulette algorithm based on crossover probability.
   iii. Mutation: randomly modify parent qDT based on mutation probability.

*Output*:
- A qDT of the best fitness.

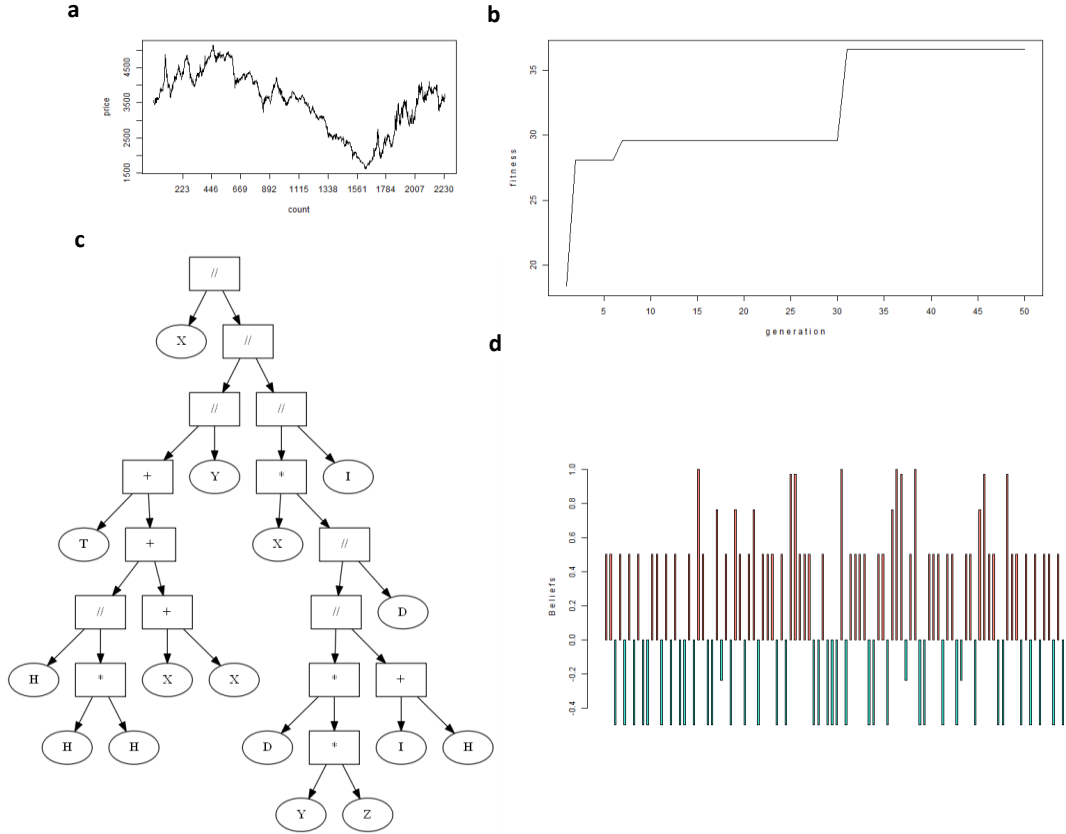

**Figure 3 Rebar trading simulated by the quantum decision tree. (a)** Price fluctuation of rebar traded on Shanghai Futures Exchange from 2009/3/27 to 2018/6/1 **(b)** Evolution of qDTs **(c)** An automatically generated qDT with the best fitness **(d)** A trader's subjective beliefs simulated by the qDT (positive red bar: people's beliefs to buy, negative green bar: people's beliefs to sell).

The k-data of rebar (transaction cycle is day) traded on the Shanghai Futures Exchange is used as the historical data for learning and optimization. Based on the quantum decision tree as in (26) (Figure 3 (c)), there are a few strategies with different subjective beliefs that the trader took (Figure 3 (d)).

$$qDT = \left( X // \left( \left( \left( T + \left( (H//(H*H)) + (X+X) \right) \right) //Y \right) // \left( \left( X * \left( \left( (D*(Y*Z))//(I+H) \right) //D \right) \right) //I \right) \right) \right) \quad (26)$$

- $X \to \hat{V} = 0.5|a_1\rangle\langle a_1| + 0.5|a_2\rangle\langle a_2|$ (50% belief to buy, 50% belief to sell)

- $Y \to \hat{V} = 0.5|a_1\rangle\langle a_1| + 0.5|a_2\rangle\langle a_2|$ (50% belief to buy, 50% belief to sell)

- $I \to \hat{V} = 0.5|a_1\rangle\langle a_1| + 0.5|a_2\rangle\langle a_2|$ (50% belief to buy, 50% belief to sell)

- $\left( T + \left( H + (X+X) \right) \right) \to \hat{V} = 0.76|a_1\rangle\langle a_1| + 0.24|a_2\rangle\langle a_2|$ (76% belief to buy, 24% belief to sell)

- $\left( T + \left( (H*H) + (X+X) \right) \right) \to \hat{V} = 0.97|a_1\rangle\langle a_1| + 0.03|a_2\rangle\langle a_2|$ (97% belief to buy, 3% belief to sell)

- $(X*D) \to \hat{V} = 0.5|a_1\rangle\langle a_1| + 0.5|a_2\rangle\langle a_2|$ (50% belief to buy, 50% belief to sell)

- $\left( X * \left( D * (Y*Z) \right) \right) \to \hat{V} = 0.5|a_1\rangle\langle a_1| + 0.5|a_2\rangle\langle a_2|$ (50% belief to buy, 50% belief to sell)

- $\left( X * (I+H) \right) \to \hat{V} = |a_1\rangle\langle a_1|$ (100% belief to buy)

## Synapse's quantum value operator hypothesis

As shown in Figure 4, neuron transmits information through synapse with the help of neurotransmitters stored in vesicles at the end of axons.[17, 18] Action potentials propagating along axons release neurotransmitters, and receptors on the postsynaptic membrane respond specifically to certain neurotransmitters to form ion channels with switching function to excite or inhibit of postsynaptic neurons. When the ion channel is opened, the incoming ions will reduce the synaptic potential difference inside and outside the cell membrane, which is called depolarization, resulting in the excitation; the outgoing ions will increase the potential difference, which is called hyperpolarization, resulting in the inhibition of neurons. While the ion channel is opened or closed, the synaptic potential changes continuously to form excitation or inhibitory synaptic inputs into the neuron body for integration, when integrated synaptic potential exceeds the threshold, the all-or-nothing action potential will be triggered. Information transmissions between neurons is to repeatedly convert digital signals into analog signals, and then restore stimulated signals into digitals, as John Von Neumann suggested[19].

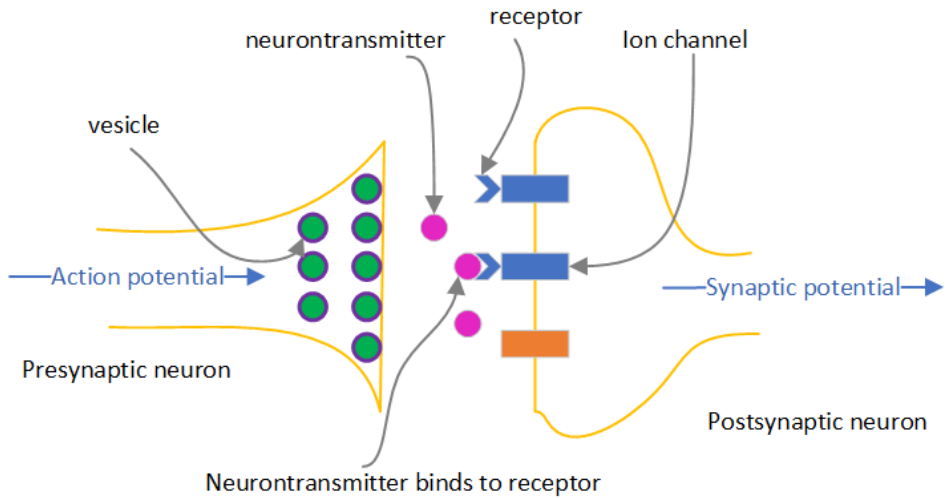

**Figure 4** Synapse structure and functions

At the neurobiological level, quantum theory was applied to reveal how the brain makes decisions. Roger Penrose and Stuart Hameroff collaborated to produce a theory known as orchestrated objective reduction[20, 21]; Henry Stapp[22] hypothesized that the information transmission at the synaptic boutons obeys the quantum mechanism; Quantum tunneling equation was used by Friedrich Beck and John Eccles[23] to describe exocytosis triggered by action potential; Mathew Fisher[24] hypothesized that phosphorus nuclear-spin may constitute qubits in the brain for information transmission; Andrei Khrennikov[25] et al. attempted to use the quantum state superposition representing action potentials to construct the quantum-like decision theory at the neuron level.

We can analogy that a single ion channel as a "trader", an ion channel has two "actions" to choose: open or close, equivalent to trader's buy or sell actions. Synapse can be compared with market; the nature states of synapse are excitation and inhibition, equivalent to market's rising or falling. Synaptic potential can be analogized to market's price and the change of synaptic potential can be compared to value of profit or loss. We can represent synapse states $|\psi\rangle$ in terms of the Hilbert state space, Hilbert strategy space $|S\rangle$ is used to represent ion channel's "strategies" which decide if open or close, Neurotransmitter combined with receptor functioned as quantum value operator can be represented as density operator $\hat{V}$ which projects an ion channel's "beliefs" onto an open or close action:

$$|\psi\rangle = c_1|q_1\rangle + c_2|q_2\rangle \tag{27}$$

$$|S\rangle = \mu_1|a_1\rangle + \mu_2|a_2\rangle \tag{28}$$

$$\hat{V} = p_1|a_1\rangle\langle a_1| + p_2|a_2\rangle\langle a_2| \tag{29}$$

Where $|q_1\rangle$ indicates a state in which the synapse is exciting and $|q_2\rangle$ indicates synapse is inhibiting; $|a_1\rangle$ refers to open ion channel and $|a_2\rangle$ refers to close ion channel; $p_1$ represents the "subjective" probability which an ion channel opened and $p_2$ represents the probability ion channel closed.

The binding of neurotransmitter and receptor as a "Von Neumann projection measurement" opens or closes an ion channel and the changed synaptic potential as observed value $V_k$. It is the natural selection that has evolved the corresponding neurotransmitter and their receptors as the fittest "quantum value operator" to guide the synapse's "decision-making". If synapse is exciting and an ion channel chooses to open, the value $V_k$ is positive and the ion channel gets an award, otherwise value $V_k$ is negative and the ion channel gets a punishment; if synapse is inhibiting and an ion channel chooses to close, the value $V_k$ is positive and the ion channel gets an award, otherwise value $V_k$ is negative and the ion channel gets a punishment. The "quantum value operator" formed by neurotransmitter and receptor evolves as described in equations (23)-(25).

## Discussion

Human beings record a large amount of data through the observation of the world. It is through the study of the recorded data that human beings gradually understand the objective world and build a simplified subjective "world model" in the brain. People make decisions by considering both the world's objectivity and the subjectivity of their beliefs. Observed value is a bridge between objective world and subjective beliefs.

We hypothesize that an uncertain objective natural state can be represented by superposition of all possible states and we also hypothesize that undecided subjective decision state can be represented by superposition of all possible actions. The information we can get regarding decision is observed value which can be represented by a quantum density operator as value operator. Value operator represents our subjective beliefs of taking actions. The process of decision is equivalent to evolution of the mixed state's density operator based on nature states, and the final decision is to perform a "quantum measurement" on the strategy state which chooses an action probabilistically and get a value.

We proposed a quantum decision theory based on quantum expected value. Instead of value function used in reinforcement learning[26], value operator is used to simulate brain's beliefs under uncertainty. Classic decision theory asks the bit question, 0 or 1? There are only two possible answers - 0 or 1 (either-or). Quantum decision theory asks the qubit question, 0 and 1? Now, the answer could have infinite possibilities (both-and). Quantum decision theory has inherent uncertainty due to superposition of quantum states and an observable result is obtained from the "collapse" of the state. A quantum value operator computes the probability of taking an action due to incomplete information and the result usually cannot be given by a definite cause, but can only be obtained probabilistically.

Descartes' dualism holds that the body is separate from the soul, until now the mind-body problem[27-31] is still a mystery. We believe that the information of the objective world is encoded in action potential, and probabilistically transformed into subjective signals at synapse. Our proposed synapse's quantum value operator hypothesis states that the combination of neurotransmitter and receptor functioned as "quantum value operator" controls the opening and closing of ion channel which to decide the firing rate of neurons. The firing pattern of the neural network expresses the decision of the brain, that is, subjective beliefs converted from the objective information.